\documentclass[pra,twocolumn,showpacs,superscriptaddress,groupedaddress]{revtex4-1}

\usepackage{amssymb,amsfonts,amsmath} 

\usepackage{graphicx} 

\DeclareMathOperator{\tr}{Tr}

\newcommand{\bb}[1]{\mathbf{#1}}

\newcommand{\be}{\begin{equation}}
\newcommand{\ee}{\end{equation}}
\newcommand{\bea}{\begin{align}}
\newcommand{\eea}{\end{align}}
\newcommand{\bmat}{\begin{bmatrix}}
\newcommand{\emat}{\end{bmatrix}}

\newcommand{\vas}{\left (}
\newcommand{\oik}{\right )}

\newcommand{\aver}[1]{\left\langle #1 \right\rangle}

\newcommand{\varder}[2]{\frac{\delta #1}{\delta #2}}
\newcommand{\varu}[3]{\frac{\delta #1}{\delta \phi(#2,#3)}}
\newcommand{\varG}[3]{\frac{\delta #1}{\delta G(#2,#3)}}
\newcommand{\unolla}[1]{\vas #1 \oik_{\phi=0}}
\newcommand{\varus}[3]{\frac{\delta #1}{\delta \phi_{#2 #3}}}

\newcommand{\rr}{\bb{r}}
\newcommand{\rry}{\bb{r}_1}
\newcommand{\rrk}{\bb{r}_2}

\newcommand{\evec}{\bb{e}}

\newcommand{\s}{\bb{s}}

\newcommand{\p}{\bb{p}}
\newcommand{\q}{\bb{q}}
\newcommand{\kk}{\bb{k}}

\newcommand{\psid}{\psi^\dagger}

\newcommand{\FF}{\mathcal{F}}

\newcommand{\taup}{\tau^\prime}
\newcommand{\taupp}{\tau^{\prime\prime}}
\newcommand{\tauintegral}[2]{\int\limits_{#1}^{#2}d\,(\tau_1-\tau_2)\,}

\newcommand{\tauintp}[2]{\int\limits_{#1}^{#2}d\,\taupp\,}

\newcommand{\patau}[1]{\frac{\partial #1}{\partial \tau}}

\newcommand{\kron}[1]{\delta_{#1}}

\begin{document}


\title{Collective modes and the speed of sound in the Fulde-Ferrell-Larkin-Ovchinnikov  state}


\author{M.O.J.\ Heikkinen} 
\affiliation{Department of Applied Physics,  
Aalto University School of Science, P.O.Box 15100, FI-00076 Aalto, FINLAND} 
\author{P. T\"orm\"a}
\email{paivi.torma@aalto.fi}
\affiliation{Department of Applied Physics,  
Aalto University School of Science, P.O.Box 15100, FI-00076 Aalto, FINLAND}




\begin{abstract}
We consider the density response of
a spin-imbalanced ultracold Fermi gas in an optical lattice in the 
Fulde-Ferrell-Larkin-Ovchinnikov (FFLO) state.
We calculate the collective mode spectrum of the system
in the generalised random phase approximation and find that though the collective modes are
damped even at zero tempererature, the damping is weak enough to have
well-defined collective modes.
We calculate the speed of sound in the gas and show that it 
is anisotropic due to the anisotropy of the FFLO pairing,
which implies an experimental signature for the FFLO state. 
\end{abstract}


\maketitle

\section{\label{intro}Introduction}

Ultracold Fermi gases are dilute systems of fermionic atoms cooled down
to temperatures where quantum statistics dominates the physics.
The unprecedented experimental possibilities of controlling and tuning
the ultracold Fermi gas systems have made them an extremely succesful tool for simulating
a broad range condensed matter phenomena \cite{blochreview,stringarireview,ketterlereview}.
To highlight the topic area of this paper, the ability to control the 
number of each atom species forming the ultracold gas
has enabled the experimental study of spin-population imbalanced fermionic superfluidity
\cite{smokingun,hulet,ketterledirect,shin2008,liao2010,nascimbene2009,nascimbene2010}.

One candidate for the theoretical description of imbalanced fermionic 
superfluids is the Fulde-Ferrell-Larkin-Ovchinnikov (FFLO) state 
\cite{FF,LO,casalbuoni2004} originally derived for superconductors
in a strong magnetic field. In the FFLO state the pairing correlations
which give rise to superfluidity occur with a finite
center of mass momentum. This leads to the fact that the FFLO state exhibits
a spatially varying order parameter.

In solid state systems, there has been progress toward 
finding experimental evidence of the FFLO state in heavy fermion systems 
\cite{Radovan2003,Bianchi2003,Kumagai2006,Correa2007} and 
also in organic superconductors \cite{Lortz2007,coniglio2011}.
In the context of ultracold gases in one-dimensional confinement,
there have been experiments \cite{liao2010} in qualitative agreement with theoretical studies on
the FFLO state. 
However, the question about the existence of the FFLO state still remains undecided. 
This subject has received considerable theoretical attention in the field of ultracold gases
and several experimental procedures to probe the FFLO state
have been suggested to complement the direct imaging of the density profile. 
For instance, the radio frequency (RF) spectroscopy of the FFLO state has 
been a subject of inquiry \cite{kinnunen2006,mizushima2007}.
In the case of 1D systems studies have been made on
collective mode properties \cite{edge2009},
double occupation modulation spectroscopy \cite{korolyuk2010}
and Josephson junction analogies \cite{hui2011}
as well as RF specroscopy \cite{reza}.
Recently, Bragg scattering and RF spectroscopy were
proposed for observing the FFLO state in quasi-1D systems \cite{yakovenko2010}.
Moreover, noise correlations have been shown to contain information about
the FFLO pairing both in 1D and in higher dimensions \cite{paananen2008,luscher2008}.

In this paper we study the density response and collective modes of the FFLO state.
While several collective mode studies exist on imbalanced Fermi gases, only few of them
consider explicitly the FFLO state \cite{edge2009,radzihovsky2011,samokhin2011}.
We concentrate on two-compononent spin-imbalanced Fermi gases at finite tempereture,
in the lowest band of a 2D or a quasi-1D optical lattice and with an on-site interaction.
The optical lattice is motivated by theoretical studies indicating 
that the lattice aids the formation of the FFLO state
as the lattice dispersion improves the overlap between the Fermi surfaces of the majority
and minority components \cite{timo2,timo3}. 
Our method for calculating the collective mode spectrum is based on the generalised 
random phase approximation (GRPA) for the linear response function of the system.
The RPA \cite{Bohm1953} is a standard tool for describing collective modes of interacting fermion systems
and it was first applied to the BCS context by Anderson \cite{Anderson1958}. The method has been used to
describe also \textit{e.g}  layered superconductors \cite{cg} and the BEC-BCS crossover \cite{belkhir1994,peng2010}.
In addition to analysing the collective mode dispersion and the speed of sound, we
also study the damping properties of the collective modes. We find the interesting
result that the the anisotropic pairing of the FFLO state leads to an anisotropy in the
speed of sound in the system. Furthermore, we study a quasi-1D optical lattice in which the tunneling in two
directions of a 3D lattice is restricted, as it has been recently suggested that a quasi-1D geometry
would provide optimal conditions for the formation of the FFLO state \cite{orso2007,hu2007,parish2007,zhao2008,donghee}.

This paper continues in the next section with an introduction of the 
Hubbard model and Green's function formalism
as well as a rederivation of the FFLO Green's function.
Section \ref{sec_response1} outlines the 
linear response problem and the Kadanoff-Baym method \cite{baym61,baym62} 
for constructing self-consistent linear response approximations.
We derive the linear response function for the FFLO state in section \ref{sec_response2}. 
After this we present our main results in section \ref{sec_results1} in which
we consider the collective mode spectrum and the speed of sound in two dimensional square optical 
lattices. In section \ref{sec_results2} we discuss a quasi-1D geometry. 
Finally, we conclude our work in section \ref{sec_conc}.

\section{The FFLO state in a lattice \label{sec_basic}}

\subsection{The theoretical framework \label{sec_basic1}}

We consider a two-component Fermi gas 
confined to the lowest energy band of a square (2D) or cubic (3D) lattice with
$N_L$ sites, and describe the system with the Hubbard model with an on-site interaction.
The Hamiltonian $H_0$ of this system is
\begin{align}
H_0 =&-\sum_{\langle \rr_1,\rr_2 \rangle,\sigma} J_\sigma(\psi_{\sigma}(\rr_1) \psid_{\sigma}(\rr_2)+\psi_\sigma(\rr_2)\psid_{\sigma}(\rr_1))\nonumber\\
&-\sum_{\rr,\sigma}\mu_\sigma \psid_{\sigma}(\rr)\psi_{\sigma}(\rr)\nonumber\\
&+\sum_{\rr}U_{12}\psid_{1}(\rr)\psid_{2}(\rr)\psi_{2}(\rr)\psi_{1}(\rr).
\end{align}
Here $\sigma\in\{1,2\}$ labels the two atomic species, 
\textit{e.q.} two hyperfine states of a fermionic atom, 
and $\psid_\sigma$ 
are the fermionic annihilation and creation operators.
(The notation $\psid_\sigma(\rr)$ is slightly more convenient for the Green's function formalism
as opposed to the more conventional notation $\hat{c}_{i,\sigma}$.)
The position vector of the lattice sites is denoted by $\rr$ and the summations run over the set of all lattice sites
with $\langle \rr_1,\rr_2 \rangle$ meaning summation over nearest neighboring sites.
Moreover, $J$ is the nearest neighbour hopping energy, $\mu_\sigma$ is the chemical potential and
$U_{12}$ is the interaction strength between the two species. 
A detailed exposition of the connection of the Hubbard model parameters with experimtal parameters
for ultracold gases can be found \textit{e.g.} in \cite{jaksch1998}.
We employ a periodic boundary condition and take the convention $\hbar=1$. 

With the assumption that the system is excited by an external perturbation $H_\phi$ of the form
\begin{equation}
H_\phi = \sum_{\sigma,\nu,\rr_1,\rr_2} \phi_{\sigma\nu}(\rr_1,\rr_2,t)\psid_{\sigma}(\rr_1)\psi_{\nu}(\rr_2), \label{Hext}
\end{equation}
the total Hamiltonian is $H=H_0+H_\phi$.
The unperturbed Hamiltonian $H_0$ is assumed time-independent, but the perturbation
$H_\phi$ may have an explicit time dependence.

In the following theoretical treatment we rely on Green's 
function techniques in the Matsubara formalism \cite{mahan,bruus2004} 
\textit{i.e.} taking time as a complex parameter, 
which allows us to deal with the finite temperature more efficiently. 
The thermodynamic average of the operator $\hat{O}$ in interaction picture
in the Matsubara formalism is defined as
\be
\aver{\hat{O}} = 
\frac{\tr \vas e^{-\beta H_0} T_\tau\vas S(0,\beta)\hat{O}\oik  
\oik }{\tr\Big( e^{-\beta H_0} S(0,\beta)\Big)}.
\ee
Here $\beta=\frac{1}{k_b T}$ with $k_b$ the Boltzmann constant and $T$ the temperature.
$T_\tau$ is the time ordering operator. The complex-time $S$-matrix is defined as
\be
S(\tau,\taup)=T_\tau\exp \left ( -\tauintp{\tau}{\taup} H_\phi(\taupp) \right ). 
\ee
The single particle Green's function is then defined as 
\be
G(1,1^\prime)= - \aver{T\vas\psi(1)\psid(1^\prime)\oik}. \label{SPGreenDef}
\ee
The shorthand notation 1 is used for the variables $\rr_1\tau_1\sigma_1$.

It is convenient to extend the range of the spin index $\sigma\in\{1,2\}$ by defining that
for $\sigma\in\{3,4\}$ one takes $\psi_\sigma=\psid_{\sigma-2}$ and $\psid_\sigma=\psi_{\sigma-2}$.
With this extension the definition for the Green's 
function above covers also the so called anomalous 
correlators in which two creation or two annihilation operators appear.
These functions are essential in describing pairing correlations in the system
on the mean field level. A similar extension is useful for $J$, $\mu$, $\phi$ and $U$;
for $\sigma_1,\sigma_2\in\{3,4\}$ one defines
$J_{\sigma_1}=-J_{\sigma_1-2}$, 
$\mu_{\sigma_1}=-\mu_{\sigma_1-2}$, 
$\phi_{\sigma_1,\sigma_2}=-\phi_{\sigma_2-2,\sigma_1-2}$ and 
$U_{\sigma_1,\sigma_2}=U_{\sigma_1-2,\sigma_2-2}$. In these definitions the choice of sign
allows to write the equations of motion in the most fluent form.

In certain expressions involving two or more field operators evaluated at the same time $\tau$,
the notations $\tau^+$ and $\tau^-$ specify the time ordering. These notations imply taking the limits
where $\tau^+\rightarrow\tau$ from the positive imaginary axis and $\tau^-\rightarrow\tau$ from the negative
imaginary axis.

The single particle Green's function follows the equation of motion
\begin{align}
&\int G_0^{-1}(1,\bar{1})G(\bar{1},1^\prime)=\nonumber\\
&\delta(1,1^\prime)+\int \phi(1,\bar{1})G(\bar{1},1^\prime)+\int \Sigma(1,\bar{1})G(\bar{1},1^\prime). \label{eom1}
\end{align}
Here the integral sign is a shorthand notation for summation over position and spin 
in addition to integration over time. 
The overbar indicates a variable of summation and integration.
Here, the potential $\phi$ appears formally as non-local in time
but only local potentials are required in the work at hand.
The inverse non-interacting single particle Green's function is
\be
G_0^{-1}(1,1^\prime) = \vas -\patau{}+ K_{\sigma_1} + \mu(1) \oik\delta(1,1^\prime), 
\ee
where $K_\sigma$ is the kinetic energy operator defined by 
$K_\sigma f(\rr) = J_\sigma\sum_{\langle \rr,\rr^\prime\rangle}(f(\rr^\prime)-f(\rr))$.
Furthermore, $\delta(1,1^\prime)$ stands for the Dirac and Kronecker 
delta for continuous and discrete variables, respectively. 
For a gereneral two-body interaction the self-energy $\Sigma$ is defined as
\be
\Sigma(1,1^\prime)
=-\int V (1,\bar{1})G_2(1,\bar{1}^-,\bar{2},\bar{1}^+)G^{-1}(\bar{2},1^\prime),
\ee
in which the two particle Green's function $G_2$ is given by
\be
G_2(1,2,3,4)=\aver{T_\tau\vas\psi(1)\psi(2)\psid(4)\psid(3)\oik}.\label{Green2}
\ee
For the on-site interaction $V(1,2)=U_{\sigma_1\sigma_2}\delta(\rr_1,\rr_2)\delta(\tau_1-\tau_2)$
the self-energy simplifies somewhat due to two trivial summations.
The equation of motion can also be inverted for $G^{-1}$ as
\be
G^{-1}(1,1^\prime)=G_0^{-1}(1,1^\prime)-\phi(1,1^\prime)-\Sigma(1,1^\prime). \label{liike}
\ee

\subsection{Imbalanced superfluid in a mean field model \label{sec_basic2}}

In this section we consider the FFLO mean field description for a spin-imbalanced Fermionic superfluid
first studied by \cite{FF,LO}. Notice that in this section the focus is on ground state
properties and the external field $H_\phi$ is not needed.

The equation of motion for the single particle Green's function involves the two particle Green's function \eqref{Green2}.
This object follows again its own equation of motion which involves the three particle Green's function,
and continuing this way, one derives an infinite set of equations known as the Martin-Schwinger hierarchy. 
In practise, one needs to decouple this hiearchy on some level 
in order to find the single particle Green's function.
We resort to a mean field approximation in equation \eqref{eom1}.
This is often referred to as the Hartree-Fock-Gor'kov approximation. In this approximation and with the notation $U=U_{12}$ the self-energy is
\begin{align}
\Sigma=& 
\delta(\rr_1\tau_1,\rr_1^\prime\tau_1^\prime)U
\left[\begin{matrix}
 -G_{22}& G_{12} & 0 &  -G_{14}\\
 G_{21} & -G_{11} & -G_{23} & 0\\
 0 & -G_{32} & -G_{44} & G_{34}\\
 -G_{41} & 0 & G_{43} & -G_{33}\\
\end{matrix}\right].
\label{hirmu}
\end{align}
All of the Green's functions appearing in $\Sigma$ have the variables 
$(\rr_1\tau_1,\rr_1\tau_1^+)$.

In a system with uniform density, the Hartree terms on the diagonal 
can be absorbed into the chemical potentials.
The Fock-exchange terms with $G_{12}$, $G_{21}$, $G_{34}$ and $G_{43}$ 
are negligible in solving for the ground state for ultracold Fermi gases, as spin-flips are
energetically highly unfavourable in experimentally relevant magnetic fields.
Finally we introduce the key element of the mean field FFLO theory. We assume that
the pairing correlations have an oscillating structure so that the
self-energy is
\begin{align}
\Sigma
=&
\delta(\rr_1\tau_1,\rr_1^\prime\tau_1^\prime)\Delta\nonumber\\
\times&
\bmat
0&0&0&   e^{2i\q\cdot\rry}\\
0&0& - e^{2i\q\cdot\rry}&0\\
0&- e^{-2i\q\cdot\rry}&0&0\\
  e^{-2i\q\cdot\rry}&0&0&0\\
\emat.
\end{align}
Here $\q$ is the FFLO pairing vector.
In the special case of $\q=0$ and $N_1=N_2$ the system
is reduced to the standard BCS description. The case with $\q=0$ and $N_1\ne N_2$ is
commonly known as the breached pair (BP) state. 

The quantity $\Delta e^{-2i\q\cdot\rry}$ is the order parameter of the FFLO state. 
In general, $\Delta$ is related
to the energy gap of pair breaking excitations.
We point out that this choice of the order parameter is not the only possibility,
and it has been shown theoretically \cite{LO,casalbuoni2004} that for instance an order parameter
of a cosine form would be energetically more favourable. However, the current choice allows
for developing the theory analytically much further, thus making the physics more transparent.

To quarantee the consistency of the mean field solution, we must have
\be
G_{32}(1,1^+)=\frac{1}{U}\Delta e^{-2i\q\cdot\rry}.\label{fflogap}
\ee
This condition is the FFLO gap equation.
In the FFLO self-energy all the nonzero elements are connected by complex conjugation or
anticommutation relations of the field operators. Therefore, one indeed has just 
one independent gap equation.

One can also fix the expected particle numbers for each atom species, $N_\sigma$, with the number equations
\be
N_{\sigma}=\sum G_{\sigma\sigma}(\bar{\rr}\tau,\bar{\rr}\tau^+).\label{fflonumber}
\ee
However, for a uniform density distribution one may write the
number equation directly in terms of the density 
(or more precisely the filling fraction) $n_\sigma=N_\sigma/N_L$.
The number equation is
\be
n_\sigma= G_{\sigma\sigma}(\rr\tau,\rr\tau^+).\label{fflonumber2}
\ee
In order to find out the values of
$\Delta$, $\mu_1$ and $\mu_2$ for any given FFLO state with pairing vector $\q$ 
and particle numbers $N_1$ and $N_2$ we need
to find a solution for the gap and number equations for the state.

In the following we derive a closed algebraic form for
the FFLO Green's function in momentum and frequency space and rewrite the gap and number
equations accordingly.

It is possible to solve the Green's function in the present approximation analytically
in the Fourier space. Here, in order to find an algebraically closed set of Fourier components
we have to pay particular attention to the broken translation invariance of the FFLO order parameter.
However, our system is still translation invariant with respect to time.
Thus, we have $G(\tau_1,\tau_2)=G(\tau_1-\tau_2)$ and we may take the 
Fourier transformation in time directly with respect to $\tau_1-\tau_2$.

We define the Fourier transformation of the Green's function as
\begin{align}
G(\p_1,\p_2,\omega)=&\sum_{\rr_1,\rr_2} \tauintegral{0}{\beta}e^{i\omega(\tau_1-\tau_2)} \nonumber\\
&
\FF(\p_1\cdot\rry) G(\rry,\rrk,\tau_1-\tau_2)\FF^\dagger(\p_2\cdot\rrk), \label{fouriertrans}
\end{align}
where the Fourier transfrom matrix $\FF$ is given by
\be
\FF(\p_1\cdot\rry)=
\bmat
e^{-i\p_1\cdot\rry} & 0 & 0 & 0\\
0 & e^{-i\p_1\cdot\rry} & 0 & 0\\
0 & 0 & e^{ i\p_1\cdot\rry} & 0\\
0 & 0 & 0 & e^{ i\p_1\cdot\rry}\\
\emat.
\ee
Here $\p_1$ and $\p_2$ are momenta and $\omega$ is a frequency (or energy, as we have
chosen the convention $\hbar=1$). The sign convention of $\FF$
has been chosen so that it agrees with the Fourier transformation of the field operators.
Due to the periodic boundary condition in complex time, also the frequency spectrum is
discrete covering the fermionic Matsubara frequencies $\omega=\frac{(2n+1)\pi}{\beta}$
where $n$ is an integer. 

The inverse Fourier transformation is 
\begin{align}
&G(\rr_1,\rr_2,\tau_1-\tau_2)=\frac{1}{\beta N_L} \sum_{\p_1,\p_2,\omega}\nonumber\\
&e^{-i\omega(\tau_1-\tau_2)}\FF^\dagger(\p_1\cdot\rry) G(\p_1,\p_2,\omega)\FF(\p_2\cdot\rrk).
\end{align}
Here the momentum summations run over the discrete momentum spectrum and the
frequency summation over the Matsubara frequencies.

We now Fourier transform the equation of motion  \eqref{eom1}  
for $G$ in the FFLO state. For brevity,
we deal first with the $\sigma_1,\sigma_2\in\{1,4\}$ block,
which has the Fourier transform
\begin{align}
&\bmat
i\omega-\xi_1(\p_1) & 0\\
0 & i\omega+\xi_2(\p_1)\\
\emat
G(\p_1,\p_2,\omega)
\nonumber\\
&=
\kron{\p_1,\p_2}I+\Delta
\bmat
0 & 1\\
1 & 0\\
\emat
G(2\q-\p_1,\p_2,\omega). \label{jepjep}
\end{align}
Here the non-interacting particle energy $\xi_\sigma(\p)$ is
\begin{equation}
\xi_\sigma(\p)=\epsilon(\p)-\mu_\sigma,
\end{equation}
in which $\epsilon(\p)$ is the lattice dispersion.

Equation \eqref{jepjep} is mixing different Fourier
components of the Green's function due to the broken translation invariance
of the order parameter.
To be more explicit, for $G_{11}$ we have
\begin{align}
&(i\omega-\xi_1(\p_1))G_{11}(\p_1,\p_2,\omega)\nonumber\\
=&\kron{\p_1,\p_2}
+\Delta G_{41}(2\q-\p_1,\p_2,\omega).
\end{align}
However, the equation is still closed.
Relabelling $\p_1$ with $2\q-\p_1$
the equation for $G_{41}(2\q-\p_1,\p_2,\omega)$ is 
\begin{align}
&(i\omega+\xi_2(2\q-\p_1))G_{41}(2\q-\p_1,\p_2,\omega)\nonumber\\
=&\Delta G_{11}(\p_1,\p_2,\omega).
\end{align}
From this pair of equations it is straightforward to solve for $G_{11}$ and $G_{41}$.
In a similar manner we find the Green's functions $G_{14}$ and $G_{44}$.
The solution is
\begin{align}
&
\left[\begin{matrix}
G_{11}(\p_1,\p_2,\omega) & G_{14}(\p_1,2\q-\p_2,\omega)\\
G_{41}(2\q-\p_1,\p_2,\omega) & G_{44}(2\q-\p_1,2\q-\p_2,\omega)\\
\end{matrix}\right]\nonumber\\
&=
\frac{\delta_{\p_1,\p_2}}{(i\omega-\xi_1(\p_1))(i\omega+\xi_2(2\q-\p_1))-\Delta^2}\nonumber\\
&\times
\left[\begin{matrix}
i\omega+\xi_2(2\q-\p_1) & \Delta \\
\Delta & i\omega-\xi_1(\p_1) \\
\end{matrix}\right].
\label{fflogreen1}
\end{align}
The solution for the Green's functions in the matrix block $\sigma_1,\sigma_2\in\{2,3\}$ 
is similar and can be written as
\begin{align}
&
\left[\begin{matrix}
G_{22}(2\q-\p_1,2\q-\p_2,\omega) & G_{23}(2\q-\p_1,\p_2,\omega)\\
G_{32}(\p_1,2\q-\p_2,\omega) & G_{33}(\p_1,\p_2,\omega)\\
\end{matrix}\right]\nonumber\\
&=
\frac{\delta_{\p_1,\p_2}}{(i\omega+\xi_1(\p_1))(i\omega-\xi_2(2\q-\p_1))-\Delta^2}\nonumber\\
&\times
\left[\begin{matrix}
i\omega+\xi_1(\p_1) & -\Delta\\
-\Delta & i\omega-\xi_2(2\q-\p_1)\\
\end{matrix}\right].
\label{fflogreen2}
\end{align}
Notice that all the other Green's functions $G_{\sigma\nu}$ are trivially zero in the
adopted approximation.

The solution can be written in a form which is easier to analyse and
is similar to the conventional form for the BCS Green's functions. 
This final step makes the
application of finite temperature Matsubara summation techniques straightforward.
One defines the well-known quasiparticle energies $E_\pm$ \cite{FF} as
\begin{align}
E_\pm(\p)&=\pm\frac{\xi_1(\p)-\xi_2(2\q-\p)}{2}\nonumber\\
&+\sqrt{\vas\frac{\xi_1(\p)+\xi_2(2\q-\p)}{2}\oik^2+\Delta^2},
\end{align}
and the coherence factors $u$ and $v$ as
\begin{align}
u(\p)=\sqrt{\frac{E_+(\p)+\xi_2(2\q-\p)}{E_+(\p) + E_-(\p)}},\\
v(\p)=\sqrt{\frac{E_-(\p)-\xi_2(2\q-\p)}{E_+(\p) + E_-(\p)}}.
\end{align}
With these definitions the FFLO Green's functions can be written 
for the block $\sigma_1,\sigma_2\in\{1,4\}$ as
\begin{align}
&
\left[\begin{matrix}
G_{11}(\p_1,\p_2,\omega) & G_{14}(\p_1,2\q-\p_2,\omega)\\
G_{41}(2\q-\p_1,\p_2,\omega) & G_{44}(2\q-\p_1,2\q-\p_2,\omega)\\
\end{matrix}\right]\nonumber\\
=&
\frac{\kron{\p_1,\p_2}}{i\omega-E_+(\p_1)}
\left[\begin{matrix}
u(\p_1)^2 & u(\p_1)v(\p_1) \\
u(\p_1)v(\p_1) & v(\p_1)^2\\
\end{matrix}\right]\nonumber\\
+&
\frac{\kron{\p_1,\p_2}}{i\omega+E_-(\p_1)}
\left[\begin{matrix}
v(\p_1)^2 & -u(\p_1)v(\p_1) \\
-u(\p_1)v(\p_1) & u(\p_1)^2\\
\end{matrix}\right]
\label{fflogreen3},
\end{align}
and for the block $\sigma_1,\sigma_2\in\{2,3\}$ as
\begin{align}
&
\left[\begin{matrix}
G_{22}(2\q-\p_1,2\q-\p_2,\omega) & G_{23}(2\q-\p_1,\p_2,\omega)\\
G_{32}(\p_1,2\q-\p_2,\omega) & G_{33}(\p_1,\p_2,\omega)\\
\end{matrix}\right]\nonumber\\
=&
\frac{\kron{\p_1,\p_2}}{i\omega-E_-(\p_1)}
\left[\begin{matrix}
u(\p_1)^2 & -u(\p_1)v(\p_1) \\
-u(\p_1)v(\p_1) & v(\p_1)^2\\
\end{matrix}\right]\nonumber\\
+&
\frac{\kron{\p_1,\p_2}}{i\omega+E_+(\p_1)}
\left[\begin{matrix}
v(\p_1)^2 & u(\p_1)v(\p_1) \\
u(\p_1)v(\p_1) & u(\p_1)^2\\
\end{matrix}\right]
\label{fflogreen4}.
\end{align}
While the normal Green's functions are diagonal in momentum space, the anomalous Green's functions are not,
reflecting the oscillatory structure of the pairing field.

Finally, we present the gap equation \eqref{fflogap} and the number equations \eqref{fflonumber} in Fourier space.
The inverse Fourier transform of $G_{32}(1,1^+)$ appearing in the gap equation is
\begin{align}
&G_{32}(\rr\tau,\rr\tau^+)=\nonumber\\
&\frac{1}{N_L}\sum_\p e^{-2i\q\cdot\rr}u(\p)v(\p)( 1 - n_F(E_+(\p)) - n_F(E_-(\p))),
\end{align}
in which the Fermi distribution $n_F$ has been obtained from the Matsubara summation
\begin{align}
n_F(E)=\frac{1}{\beta}\sum_\omega \frac{e^{i\omega(\tau^+-\tau)}}{i\omega-E}.
\end{align}
Thus, the gap equation is
\begin{align}
\frac{\Delta}{U}=\frac{1}{N_L}\sum_\p u(\p)v(\p)( 1 - n_F(E_+(\p)) - n_F(E_-(\p)) ). \label{fflogap2}
\end{align}
Similarly the number equations in terms of the filling fraction are
\begin{align}
&n_1=G_{11}(\rr\tau,\rr\tau^+)=\nonumber\\
&\frac{1}{N_L}\sum_\p u(\p)^2 n_F(E_+(\p))+v(\p)^2 (1-n_F(E_-(\p))),
\end{align}
and
\begin{align}
&n_2=G_{22}(\rr\tau,\rr\tau^+)=\nonumber\\
&\frac{1}{N_L}\sum_\p u(\p)^2 n_F(E_-(\p))+v(\p)^2(1- n_F(E_+(\p))).
\end{align}

\section{Density response in the FFLO state \label{sec_response}}

\subsection{Linear response theory \label{sec_response1}}

Having established a ground state description for the imbalanced superfluid, 
let us turn to study density fluctuations caused by potentials that couple to
the particle density. The Hamiltonian $H_\phi$ for such external potentials is of the form
\begin{align}
H_\phi&=\int\,\phi_1(\bar{\rr} t)\psid_1(\bar{\rr})\psi_1(\bar{\rr})\nonumber\\
&+\int\,\phi_2(\bar{\rr} t)\psid_2(\bar{\rr})\psi_2(\bar{\rr}).
\end{align}
Now, if and when solving the system with $H=H_0+H_\phi$ directly is not feasible, progress can be made
by assuming that $H_\phi$ is a small perturbation. To find out the effect of $H_\phi$ for instance on 
the density of atom species $\sigma=1$
one can then write $n_{\sigma_1}(\rr t)$ as a variational series with respect to $\phi$
and truncate this series to the first order, obtainging
\begin{align}
n_1(\rr t)&=\unolla{n_1(\rr t)}
+\int \phi_1(\bar{\rr}\bar{t})\unolla{\varder{n_1(\rr t)}{\phi_1(\bar{\rr}\bar{t})}}\nonumber\\
&+\int \phi_2(\bar{\rr}\bar{t})\unolla{\varder{n_1(\rr t)}{\phi_2(\bar{\rr}\bar{t})}}. \label{basicresponse}
\end{align}

Let us continue in the Matsubara formalism.
The concept above can be generalised for any Green's function 
by writing it as a variational series to the first order with respect to $\phi(1,2)$ so that
\begin{align}
G(1,1^\prime)&=\unolla{G(1,1^\prime)}+ \int \phi({\bar{2}},{\bar{3}})\unolla{\varu{G(1,1^\prime)}{\bar{2}}{\bar{3}}}. 
\end{align}
Notice again, that the overbar indicates summation/integration over position, time and spin.
The variational derivative in the equation above defines the linear response function
\be
L(12,1^\prime2^\prime)=\unolla{\varu{G(1,1^\prime)}{2^\prime}{2}}.\label{defL}
\ee
For example the density is given by $n_{\sigma_1}(\rry\tau_1)=G(1,1^+)$ 
so $L(12,1^+2^\prime)$ would give the density response function.

The linear response function carries 
information about the excited states of the unperturbed system.
If one is able to solve the linear response function
one can then extract from it the excitation spectrum of the system. 
For instance the collective density modes appear as simple
poles of the density response function in frequency space.

Following Kadanoff and Baym \cite{baym61,baym62} 
one can derive an equation for the linear response
function from the equation of motion \eqref{liike}. 
To outline the derivation of their result briefly, let us begin by taking the variational derivative of 
the identity $\int GG^{-1}=\delta$, which yields
\begin{equation}
\varu{G(1,1^\prime)}{2^\prime}{2} =  - \int G (1,\bar{3} ) \varu{G^{-1}(\bar{3},\bar{4})}{2^\prime}{2} G (\bar{4},1^\prime).
\end{equation}
Now, inserting $G^{-1}$ from (\ref{liike}) into the previous equation,
evaluating the expression at $\phi=0$ and identifying $L$ we get
\begin{align}
&L(12,1^\prime 2^\prime)  = \nonumber\\
&\int \unolla{ G (1,\bar{3} ) \vas \varu{\phi(\bar{3},\bar{4})}{2^\prime}{2} 
+ \varu{\Sigma(\bar{3},\bar{4})}{2^\prime}{2} \oik G(\bar{4},1^\prime)}. \label{Leq}
\end{align}
Since the self-energy does not depend explicitly on the external perturbation, 
the chain rule of differentiation gives
\begin{align}
\varu{\Sigma(3,4)}{2^\prime}{2} 
&= \int \varG{\Sigma(3,4)}{\bar{5}}{\bar{6}} \varu{G(\bar{5},\bar{6})}{2^\prime}{2}\nonumber\\
&= \int \varG{\Sigma(3,4)}{\bar{5}}{\bar{6}} L(\bar{5}2,\bar{6}2^\prime).
\end{align}
Therefore, one may write equation \eqref{Leq} explicitly as an integral equation for the linear response function
\begin{align}
&L(12,1^\prime 2^\prime) = \nonumber\\
&\int G (1,\bar{3} )_{\phi=0}  G(\bar{4},1^\prime)_{\phi=0} \unolla{\varu{\phi(\bar{3},\bar{4})}{2^\prime}{2}}  
\nonumber\\
+& \int G (1,\bar{3} )_{\phi=0}  G(\bar{4},1^\prime)_{\phi=0}   
\unolla{\varG{\Sigma(\bar{3},\bar{4})}{\bar{5}}{\bar{6}}} L(\bar{5} 2,\bar{6} 2^\prime), \label{Leqq}
\end{align}
which is the result of \cite{baym61,baym62}. 
This equation for the response function quarantees a 
self-consistent theory in the sense that the linear response function
obeys the same conservation laws as does the single particle Green's function.
From this point on, we leave out the notation $\phi=0$ as in the following all of the
variational derivatives are evaluated at $\phi=0$. Notice that in the equation above
there is a trivial variational derivative
\begin{align}
\varu{\phi({3},{4})}{1}{2}=\delta(\rr_3\tau_3-\rry\tau_1)\delta(\rr_4\tau_4-\rr_2\tau_2)\hat{\delta}(\sigma_3\sigma_4,\sigma_1\sigma_2),
\end{align}
where we have the notation
\begin{align}
&\hat{\delta}(\sigma_3\sigma_4,\sigma_1\sigma_2)=\nonumber\\
&\delta(\sigma_3,\sigma_1)\delta(\sigma_4,\sigma_2)-\delta(\sigma_4,\sigma_1\pm 2)\delta(\sigma_3,\sigma_2\pm 2).
\end{align}
The second term in this definition owes to the fact that for the extended index 
$\sigma_1,\sigma_2\in\{3,4\}$ we have $\phi_{\sigma_1,\sigma_2}=-\phi_{\sigma_2-2,\sigma_1-2}$.

\subsection{Derivation of the FFLO density response function \label{sec_response2}}

We derive the FFLO density response function in this section.
The Kadanoff-Baym method applied to the Hartree-Fock-Gor'kov approximation \eqref{hirmu}
leads to the generalised random phase approximation (GRPA).
We now extend this to the case of a FFLO state.
In the density response problem, several simplifications to the
general equation for the linear response function are apparent.
First of all, it is sufficient to consider a local perturbation of the form
$\phi(1,2)=\phi_{\sigma_1\sigma_2}(\rry\tau_1)\delta(\rry\tau_1-\rrk\tau_2)$.
Furthermore, the density operator itself as well as all 
the Green's functions in the self-energy are local.
Thus, the density response function of which we are interested in contains only 
terms with $\rry=\rry^\prime$, $\tau_1=\tau_1^\prime$ and $\rrk=\rrk^\prime$, 
$\tau_2=\tau_2^\prime$. Moreover, due to the time translation invariance
this response function depends only on the time difference $\tau_1-\tau_2$.
We then find it suitable for our purposes to define the notation
\begin{align}
&L_{\sigma_1\sigma_2\sigma_1^\prime\sigma_2^\prime} (\rry,\rrk,\tau_1-\tau_2) = \nonumber\\
&\delta(\rry\tau_1,\rry^\prime\tau_1^\prime)\delta(\rrk\tau_2,\rrk^\prime\tau_2^\prime)L(12,1^\prime 2^\prime).
\end{align}
The density response functions $\frac{\delta n_1}{\delta\phi_1}$ and $\frac{\delta n_1}{\delta\phi_2}$
in equation \eqref{basicresponse} correspond to $L_{1111}$ and $L_{1212}$, respectively. 

Equation \eqref{Leq} now reads
\begin{align}
&L_{\sigma_1\sigma_2\sigma_1^\prime\sigma_2^\prime}(\rry , \rrk, \tau_1 - \tau_2) 
= \nonumber\\ 
&\sum
G_{\sigma_1\bar{\sigma}_3}  (\rry , \rrk,\tau_1-\tau_2) \nonumber\\
&\times G_{\bar{\sigma}_{4}\sigma_1^\prime} (\rrk,\rry,\tau_2-\tau_1)  
\hat{\delta}(\bar{\sigma}_3\bar{\sigma}_4,\sigma_2^\prime\sigma_2) \nonumber\\
&+ 
\int G_{\sigma_1\bar{\sigma}_{3}}  (\rry , \bar{\rr}_3,\tau_1-\bar{\tau}_3) \nonumber\\
&\times
G_{\bar{\sigma}_{4}\sigma_1^\prime} (\bar{\rr}_3,\rry,\bar{\tau}_3-\tau_1) 
\varus{\Sigma_{\bar{\sigma}_3 \bar{\sigma}_4}}{\sigma_2^\prime}{\sigma_2}
(\bar{\rr}_3,{\rr}_2,\bar{\tau}_3-{\tau}_2). \label{ruma}
\end{align}

Let us now write down explicitly the equation for $L_{1111}$ to show how to proceed with the solution. 
Since many of the FFLO Green's functions are identically zero, 
just one term remains from the direct coupling to the external 
perturbation which is the first term in equation \eqref{ruma}. 
Similarly, only four non-zero terms arise from the second term in equation \eqref{ruma}. 
\begin{widetext}
\begin{align}
L_{1111}(\rry , \rrk, \tau_1 - \tau_2)  
&= 
G_{11}(\rry , \rrk,\tau_1-\tau_2) G_{11}(\rrk,\rry,\tau_2-\tau_1)  \nonumber\\
&+ U\int 
G_{14}(\rry , \bar{\rr}_3,\tau_1-\bar{\tau}_3) 
G_{41}(\bar{\rr}_3,\rry,\bar{\tau}_3-\tau_1)  
L_{1111}(\bar{\rr}_3,{\rr}_2,\bar{\tau}_3-{\tau}_2)\nonumber\\
&-U\int 
G_{11}(\rry , \bar{\rr}_3,\tau_1-\bar{\tau}_3) 
G_{41}(\bar{\rr}_3,\rry,\bar{\tau}_3-\tau_1)  
L_{1141}(\bar{\rr}_3,{\rr}_2,\bar{\tau}_3-{\tau}_2)\nonumber\\
&-U\int 
G_{14}(\rry , \bar{\rr}_3,\tau_1-\bar{\tau}_3) 
G_{11}(\bar{\rr}_3,\rry,\bar{\tau}_3-\tau_1)  
L_{4111}(\bar{\rr}_3,{\rr}_2,\bar{\tau}_3-{\tau}_2)\nonumber\\
&+ U\int 
G_{11}(\rry , \bar{\rr}_3,\tau_1-\bar{\tau}_3) 
G_{11}(\bar{\rr}_3,\rry,\bar{\tau}_3-\tau_1)  
L_{4141}(\bar{\rr}_3,{\rr}_2,\bar{\tau}_3-{\tau}_2). \label{L1111}
\end{align}
\end{widetext}
In forming this equation we have used the identities $L_{3131}=-L_{1111}$ and $L_{2121}=-L_{4141}$.
This substitution owes to a more general identity: Noting 
the definitions \eqref{defL} of the response function and 
\eqref{SPGreenDef} of the Green's function one concludes based on 
equal time anticommutation relations that
\begin{align}
L_{\sigma,\alpha,\nu,\beta}(\rry , \rrk, \tau) 
&=-L_{\sigma+2,\alpha,\nu+2,\beta}(\rry , \rrk, \tau) , \nonumber\\
L_{\sigma+2,\alpha,\nu,\beta}(\rry , \rrk, \tau) 
&=-L_{\sigma,\alpha,\nu+2,\beta}(\rry , \rrk, \tau). 
\label{manipulation}
\end{align}

As with the Green's function, the integral equation 
for $L_{1111}$ \eqref{L1111} can be
cast into an algebraic equation in Fourier space. 
Notice that for instance in spherically symmetric harmonic trapping geometries
one can make similar progress with the choice of 
the harmonic oscillator states as basis functions
\cite{annanpapru}.
We use the same Fourier transformation convention for 
$L_{\sigma,1,\nu,1}=\frac{\delta G_{\sigma\nu}}{\delta \phi_{11}}$ 
as we defined for $G_{\sigma\nu}$ in \eqref{fouriertrans}
\textit{i.e.} the sign convention is given by $\sigma$ and $\nu$. 
The Fourier transformation yields for $L_{1111}$, \textit{i.e.} $\frac{\delta n_1}{\delta \phi_{11}}$, the equation
\begin{align}
&L_{1111}(\p_1 , \p_2 , \omega)  
=\kron{\p_1,\p_2}\Pi_{1111}(\p_1,\omega) \nonumber\\
+& U\Pi_{1441}(\p_1,\omega)L_{1111}(\p_1 , \p_2 ,\omega)\nonumber\\
-&U\Pi_{1141}(\p_1,\omega)L_{1141}(2\q+\p_1 , -\p_2 , \omega)\nonumber\\
-&U\Pi_{1411}(\p_1,\omega)L_{4111}(2\q-\p_1 , \p_2 , \omega)\nonumber\\
+&U\Pi_{1111}(\p_1,\omega)L_{4141}(-\p_1 , -\p_2 , \omega). \label{L1111b}
\end{align}
Here we have the notation
\begin{align}
&\Pi_{\sigma_1\sigma_2\sigma_3\sigma_4} (\p,\omega)\nonumber\\
=&\frac{1}{\beta N_L}
\sum_{\s,\chi} G_{\sigma_1\sigma_2} (\lambda_{\sigma_1}(\p+\s),\lambda_{\sigma_2}(\p+\s),\chi+\omega) \nonumber\\
\times& G_{\sigma_3\sigma_4} (\lambda_{\sigma_3}(\s),\lambda_{\sigma_4}(\s),\chi), \label{definepi}
\end{align}
where $\lambda_\sigma(\p)$ is defined so that
\begin{align}
&\lambda_{\sigma}(\p)=\p, \qquad\qquad\,\,
\sigma\in\{1,2\},\nonumber\\
&\lambda_{\sigma}(\p)=2\q-\p, \qquad \sigma\in\{3,4\}.
\end{align}
Equation \eqref{L1111} is rather analogous to the previously solved
equation for the FFLO Green's function. We see that we need to construct equations
also for $L_{1141}$, $L_{4111}$ and $L_{4141}$. Using again the symmetry property
\eqref{manipulation} for these equations we find out that no other linear response
functions enter the equation. The final task is to 
identify those Fourier components which form a closed equation.
With this rationale one arrives at the following matrix equation
\be
M^{(1)} (\p_1) L^{(1)} (\p_1,\p_2) = \kron{\p_1,\p_2}\Pi^{(1)}(\p_1). \label{SPeq}
\ee
Here the vector of linear response functions $L^{(1)}$ is defined as
\be
L^{(1)}(\p_1,\p_2)=
\bmat L_{1111}(\p_1,\p_2)\\ 
L_{1141}(2\q+\p_1,-\p_2)\\
L_{4111}(2\q-\p_1,\p_2)\\
L_{4141}(-\p_1,-\p_2) 
\emat.
\ee
On the right hand side $\Pi^{(1)}$ contains the terms from the direct coupling to the external perturbation
\be
\Pi^{(1)}(\p_1)
=
\bmat \Pi_{1111}(\p_1)\\ 
\Pi_{1114}(\p_1) \\
\Pi_{4111}(\p_1) \\
\Pi_{4114}(\p_1) \emat.
\ee
The coefficient matrix $M^{(1)}$ is given by
\begin{widetext}
\be
M^{(1)}(\p_1)=I+U
\bmat
-\Pi_{1441}(\p_1) &\Pi_{1141}(\p_1) &\Pi_{1411}(\p_1) &-\Pi_{1111}(\p_1) \\
-\Pi_{1444}(\p_1) &\Pi_{1144}(\p_1) &\Pi_{1414}(\p_1) &-\Pi_{1114}(\p_1) \\
-\Pi_{4441}(\p_1) &\Pi_{4141}(\p_1) &\Pi_{4411}(\p_1) &-\Pi_{4111}(\p_1) \\
-\Pi_{4444}(\p_1) &\Pi_{4144}(\p_1) &\Pi_{4414}(\p_1) &-\Pi_{4114}(\p_1)  \label{Ms}
\emat.
\ee
\end{widetext}

One derives similarly for $L_{1212}$, \textit{i.e.} $\frac{\delta n_1}{\delta \phi_2}$, the equation
\be
M^{(2)} (\p_1) L^{(2)} (\p_1,\p_2) = -\kron{\p_1,\p_2}\Pi^{(2)}(\p_1), \label{SPeq2}
\ee 
in which $L^{(2)}$ stands for
\be
L^{(2)}(\p_1,\p_2)=
\bmat L_{1212}(\p_1,\p_2)\\ 
L_{1242}(2\q+\p_1,-\p_2)\\
L_{4212}(2\q-\p_1,\p_2)\\
L_{4242}(-\p_1,-\p_2) 
\emat,
\ee
and $\Pi^{(2)}$ is
\be
\Pi^{(2)}(\p_1)
=
\bmat \Pi_{1441}(\p_1)\\ 
\Pi_{1444}(\p_1) \\
\Pi_{4441}(\p_1) \\
\Pi_{4444}(\p_1) \emat.
\ee
The coefficient matrix $M^{(2)}$ turns out to be the same as $M^{(1)}$.

In our theoretical treatment we have an obvious symmetry with respect to interchanging
indices 1 and 2 and indices 3 and 4. Therefore, we obtain the equation
for the density response functions $L_{2222}$ and $L_{2121}$ directly from the results above.

The frequency summation in the definition of $\Pi$, equation \eqref{definepi}, can be handled analytically.
One applies the identity
\begin{align}
\frac{1}{\beta}\sum_\chi \frac{1}{i(\omega+\chi)-E_1}\cdot\frac{1}{i\chi-E_2}=\frac{ n_F(E_1) - n_F(E_2) }{ E_1-E_2-i\omega }
\end{align}
to the four cross terms that arise when one inserts the 
Fourier transformed Green's functions \eqref{fflogreen3} and \eqref{fflogreen4}
to equation \eqref{definepi}. 
The momentum summation in \eqref{definepi} needs to be carried out numerically after which one simply inverts the
matrix equations presented above to obtain the Matsubara, \textit{i.e.} imaginary frequency, response function.
From this, the real frequency retarded linear response function is obtained by means of analytical continuation.

\section{Results \label{sec_results}}

\subsection{2D square lattice \label{sec_results1}}

In the following results we consider a system in a two dimensional square optical lattice with 
lattice constant $d$ and $N_L=N_x N_y=40000$ lattice sites with
$N_x=N_y=200$ lattice sites in each direction. We assume that the perturbation potential is the
same for both atom species, \textit{i.e.} $\phi_1=\phi_2$ in which case it is most natural 
to study the density response function
$\chi_1(\kk,\omega)=L_{1111}(\kk,\omega)+L_{1212}(\kk,\omega)$ where $L_{1111}$ 
and $L_{1212}$ are the responses of density $n_1$ to
potentials $\phi_1$ and $\phi_2$, respectively. 
In the following, the parameters are chosen so that $n_1$ is the density of the majority component.
Similar conclusions hold also for the response of the minority component.
The assumption of $\phi_1=\phi_2$ is not a crucial one as the 
collective mode dispersion is the same for all choices of these potentials.
Moreover, we assume that the FFLO vector $\q$ is directed along the $x$-axis.

We solve equations \eqref{SPeq} and \eqref{SPeq2} numerically for imaginary frequencies. 
We then use a Pad\'{e} approximant \cite{lee1996} to carry out the analytical 
continuation and obtain the response function for complex frequencies.

\begin{figure}
\includegraphics[width=0.5\textwidth]{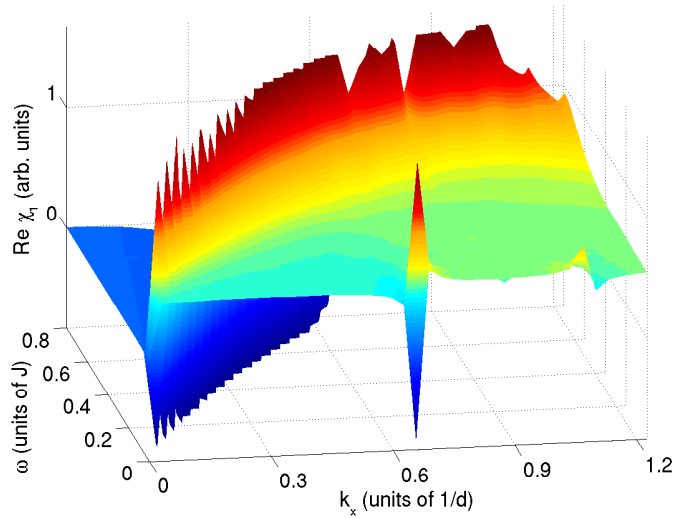} 
\caption{(Color online) The real part of the density response $\chi_1(\kk,\omega)$ for typical parameters
$\mu_1= 3.5 J$, $\mu_2= 2.5 J$, $T= 0.07 J$, $U= -3.0 J$, $\Delta= 0.27 J$ and $q= 12\pi/(dN_x)$
with $\kk$ parallel to $\q$. The collective mode is seen as a clear divergent behaviour of 
the density response.
The lone peak on the $k_x$ axis is a numerical instability.}
\label{fig:tyyppiesim} 
\end{figure}  

In Figure \ref{fig:tyyppiesim} we plot the real part of the density response function 
for typical parameters as a function of the wave vector $\kk$ and the real frequency $\omega$. 
In this figure we have chosen $\kk$ parallel to $\q$.
The collective density mode appears as nearly diverging feature of the density response at particular 
a wave vector $\kk$ and frequency $\omega$ 
Towards higher wave vectors we see a 
typical broadening of this feature owing to the increase of damping. 
The actual eigenfrequency of the mode is complex
with the frequency $\omega-i\gamma$ where $\gamma$ is the damping rate. 
For the small damping rates the response has a 
very sharp jump also as a function of real frequencies.
The collective mode is gapless with a linear dispersion for small $\kk$. 
It corresponds to the Anderson-Bogoliubov phonon of the BCS state
of a neutral Fermi gas \cite{Anderson1958}. 

\begin{figure}
\includegraphics[width=0.5\textwidth]{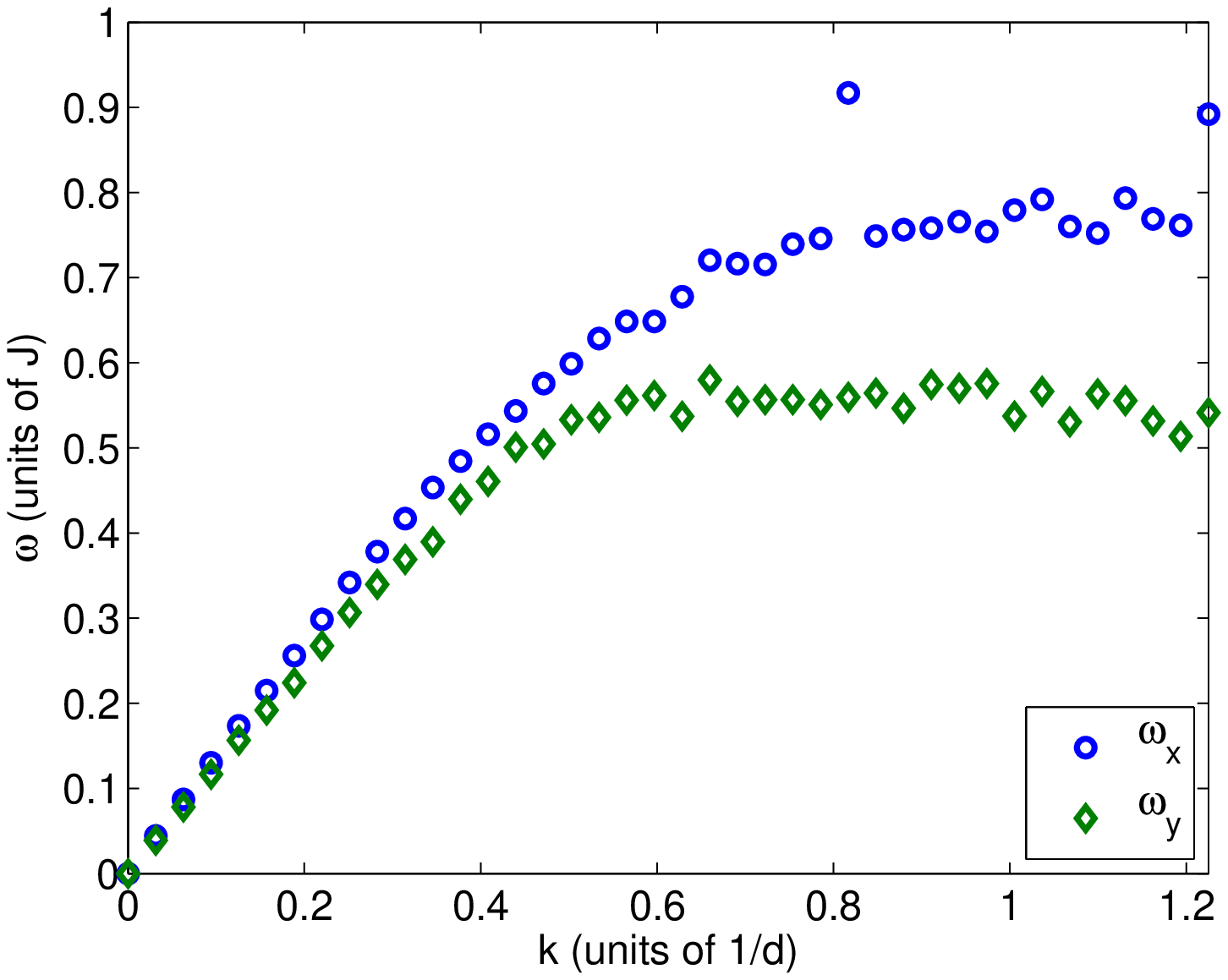} 
\includegraphics[width=0.5\textwidth]{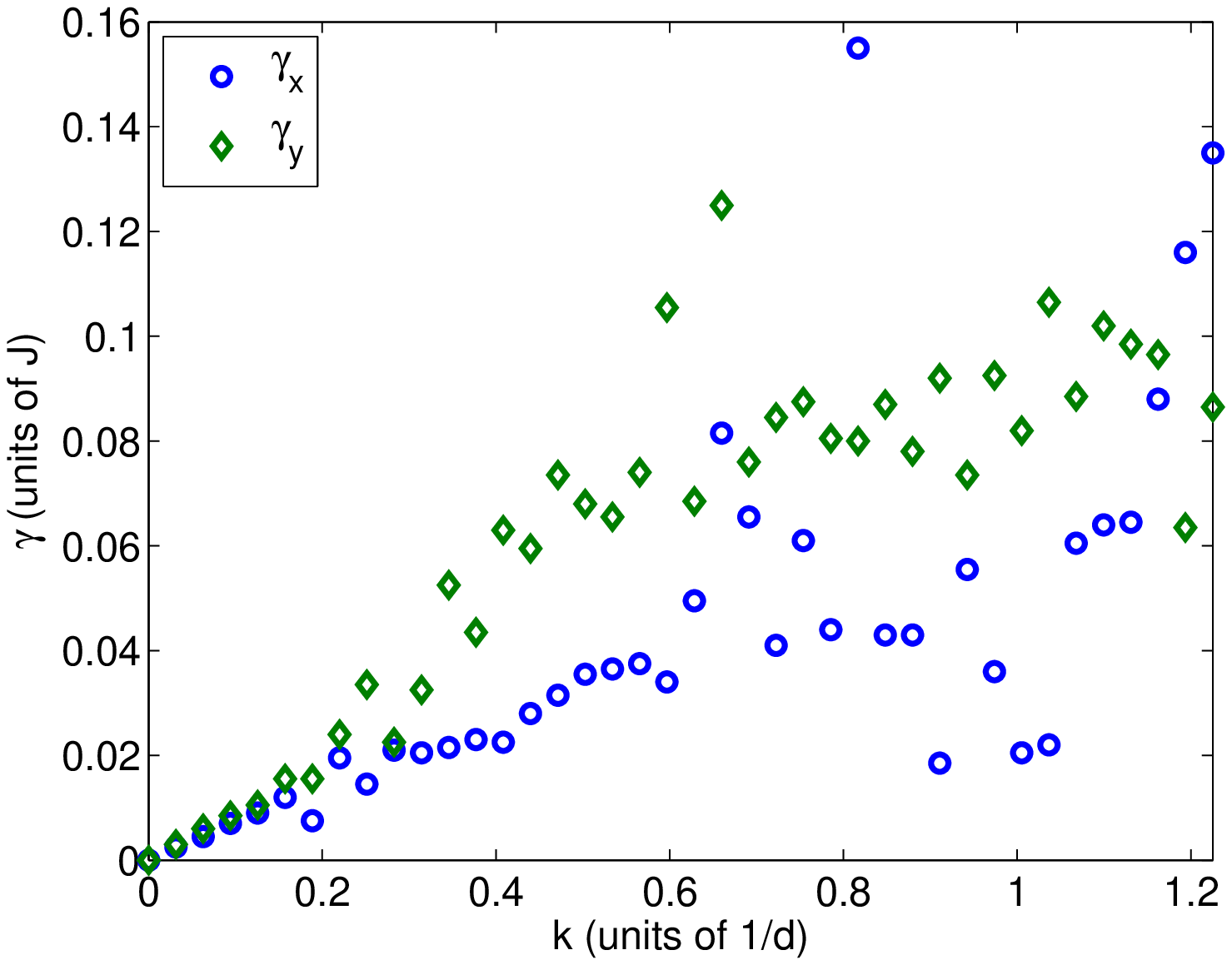} 
\caption{(Color online) The dispersion relation $\omega(\kk)$ and 
the damping rate $\gamma(\kk)$ calculated for
$\mu_1= 3.5 J$, $\mu_2=2.5J$, $T=0.07J$, $U=3.0 J$, $\Delta= 0.27 J$ 
and $q=12\pi/(dN_x)$ along the $x$-axis (circles) and $y$-axis (diamonds).
The FFLO vector $\q$ is directed along the $x$-axis and this anisotropy
creates a clear difference between the two dispersions and damping rates.} 
\label{fig:dispersiot} 
\end{figure}  

We solve the dispersion relation $\omega(\kk)$ and the damping rate $\gamma(\kk)$
by solving the poles of the response function $\chi_1$.
We plot $\omega(\kk)$ and $\gamma(\kk)$ in Figure \ref{fig:dispersiot}
for wave vectors along the $x$- and $y$-axes \textit{i.e.} parallel and perpendicular
to the FFLO vector $\q$. Notice that the numerical method
produces several instabilities in the damping rate at higher wave vectors while the real
frequency $\omega$ is far more stable.
The speed of sound to the direction of the $x$-axis, $c_x$, can be obtained from the dispersion by the definition
\begin{equation}
c_x=\left.\frac{d \omega(k \evec_x)}{d k}\right|_{k\rightarrow 0},
\end{equation}
where $\evec_x$ is the unit vector in $x$-direction. One defines $c_y$ similarly.
For the dispersions presented in Figure \ref{fig:dispersiot} we obtain $c_x= 1.39 Jd$ and $c_y= 1.24 Jd$.
We  observe that the finite FFLO vector causes a clear difference between the parallel
and perpendicular (w.r.t. $\q$) speeds of sound. In this case the relative difference
in the sound propagation is $c_{x}/c_{\mathrm{y}}-1= 12 \%$.
This observation suggests immediately an experiment in which one creates a local
density perturbation in the system and then monitors the propagation of this
perturbation to collect information about the FFLO state. Such an experiment would create a rather
remarkable contrast with \textit{e.g.} a breached pairing state, which is isotropic with $c_x=c_y$.

\begin{figure}
\includegraphics[width=0.5\textwidth]{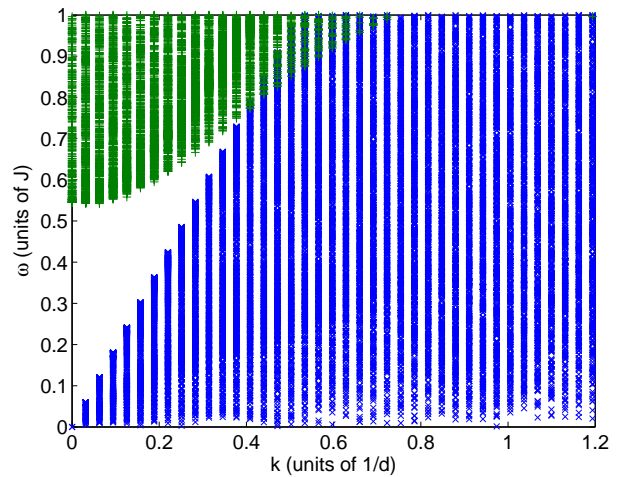} 
\caption{(Color online) An illustration of the quasiparticle transitions 
for a system with 
$\mu_1= 3.5 J$, $\mu_2= 2.5 J$, $\Delta= 0.27 J$ 
and $q= 12\pi/(dN_x)$ along the $x$-axis, corresponding to a zero temperature 
Fermi distribution.
A cross denotes a transition of a single quasiparticle 
and a plus a pair breaking which involves the creation of two quasiparticles.
Notice that the figure shows only the positions of 
the quasiparticle excitations, not their relative weights.
The small void regions along the $k_x$ axis owe to the finite system size.}
\label{fig:quasi2d} 
\end{figure}

Turning back to analyse the damping rates presented in Figure \ref{fig:dispersiot}, we
find that the damping rate is only a fraction of the mode frequency, with $\gamma_x/\omega_x= 5\%$
and $\gamma_y/\omega_y= 7\%$ for the lowest frequencies. With this observation we conclude that the
collective modes can be considered as well-defined elementary excitations of the FFLO state.
An important aspect here is that the damping is fundamentally not an effect caused by the finite
temperature. In contrast to the BCS state, the FFLO state contains also at zero temperature 
unpaired quasiparticles which cause a finite lifitime for the collective modes. In Figure 
\ref{fig:quasi2d} we illustrate the locations of the quasiparticle channels for the system of
Figures \ref{fig:tyyppiesim} and \ref{fig:dispersiot}. Indeed, the collective mode dispersion
lies deep within the region containing quasiparticle transitions. 
To compare, the pair breaking excitations are present also in a BCS state of a neutral Fermi gas, 
but at zero temperature the quasiparticle transitions
are completely absent and the Anderson-Bogoliubov phonon is thus undamped.

\begin{figure}
\includegraphics[width=0.5\textwidth]{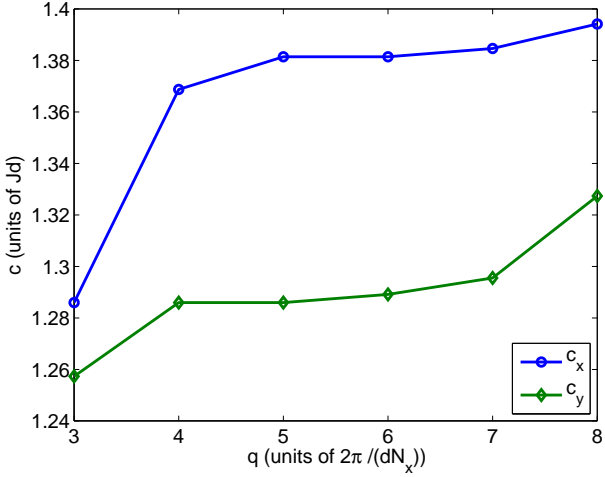} 
\caption{(Color online) The speed of sound parallel ($c_x$, circles) and perpendicular ($c_y$, diamonds)
to the FFLO vector $\q=q\evec_x$ as a function of length of the FFLO vector $q$ for a
system with $\mu_1= 3.4 J$, $\mu_2= 2.6 J$, $T= 0.1 J$, $U= -2.8 J$. 
The gap $\Delta$ is solved from the gap equation \eqref{fflogap} for each $q$.}
\label{fig:qvar_gconst} 
\end{figure}

In Figure \ref{fig:qvar_gconst} we study the speed of sound 
as a function of the length of the FFLO vector $q$ when 
the other system parameters are held constant and the gap is solved from the
gap equation \eqref{fflogap} for each $q$. 
In the range $qdN_x/(2\pi)=4\ldots 8$ we find an anisotropy of 5 to 7 percent.
The figure also indicates that the speed of sound tends to increase with $q$. This is caused by the fact
that the gap $\Delta$ decreases when $q$ increases, which in turn causes  the increase in the speed of sound.
The two smallest possible FFLO vectors have been left out, since they lead to a pair breaking gap which is larger
than the chemical potential difference $2\Delta>|\mu_1-\mu_2|$ and therefore the state is not physically relevant.

\begin{figure}
\includegraphics[width=0.5\textwidth]{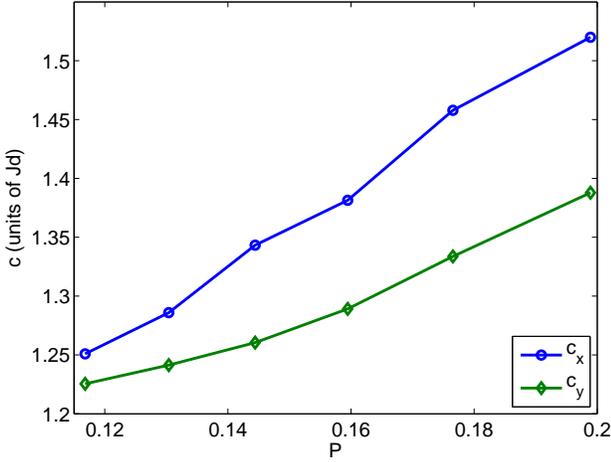} 
\caption{(Color online) The speed of sound parallel ($c_x$, circles) and perpendicular ($c_y$, diamonds)
to the FFLO vector $\q=q\evec_x$ as a function of the polarisation 
$P=(N_1-N_2)/(N_1+N_2)$. Here we vary the chemical potential 
difference $\delta\mu=\mu_1-\mu_2$ to vary the polarisation
while holding $(\mu_1+\mu_2)/2$ constant at $3.0J$. The other parameters for the system are 
$T= 0.1 J$, $U= -2.8 J$, $q=12\pi/(dN_x)$.
The gap $\Delta$ is solved from the gap equation \eqref{fflogap} for each set of parameters.}
\label{fig:muvar_gconst} 
\end{figure}

We then turn to examine the effect of the polarisation $P=(N_1-N_2)/(N_1+N_2)$ 
on the speed of sound in Figure \ref{fig:muvar_gconst}.
Technically, we vary the chemical potential difference while 
holding the average chemical potential $(\mu_1+\mu_2)/2$ and other system parameters constant.
We find that the anisotropy in the sound propagation increases from $2\%$ to $10\%$ when the polarisation
increases to the value $P=0.2$.
Also the speed of sound itself increases with the polarisation, and the explanation is analogous
to the discussion of the FFLO vector above: while holding the other system parameters fixed, the
gap $\Delta$ decreases with the chemical potential difference, or the polarisation. 
In the range $P<0.11$ the pair breaking gap is larger than the chemical potential 
difference, and the FFLO ansatz does not produce a physically relevant state.

\begin{figure}
\includegraphics[width=0.5\textwidth]{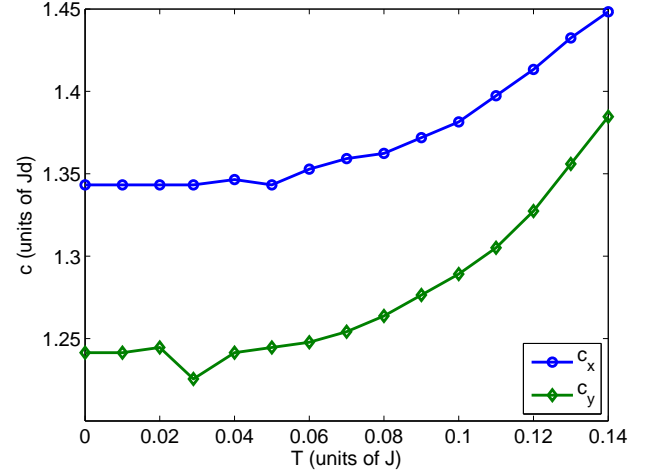}
\includegraphics[width=0.5\textwidth]{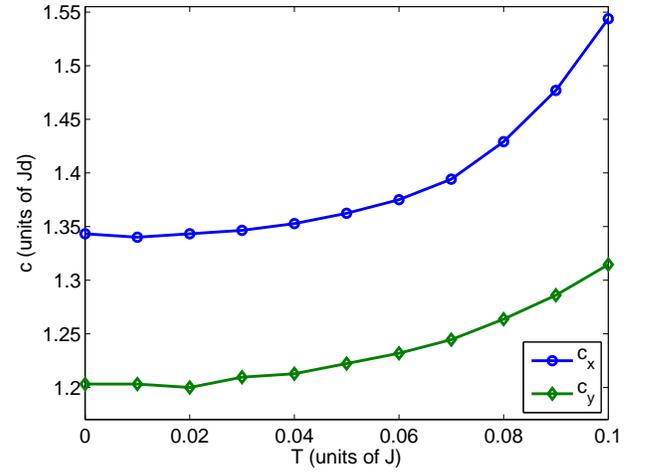}
\caption{(Color online) 
The speed of sound parallel ($c_x$, circles) and perpendicular ($c_y$, diamonds) to the FFLO vector $\q=q\evec_x$ 
as a function of the temperature $T$.
Here the system parameters are in the figure above $\mu_1= 3.4 J$, $\mu_2= 2.6 J$, $U= -2.8 J$ and $q=12\pi/(dN_x)$ 
and for the figure below $\mu_1= 3.5 J$, $\mu_2= 2.5 J$, $U= -3.0 J$ and $q=12\pi /(dN_x)$.
Note that the critical temperature when holding these parameters constant is $T_c=0.15J$ (above) and $T_c=0.12J$ (below).
}
\label{fig:temperature} 
\end{figure}

We examine the tempereture dependence of the speed of sound in Figure \ref{fig:temperature}. 
Firstly, we stress that while
the temperature does affect the dispersion relation, it always remains true that in the small
wavelength limit the collective mode is massless and the dispersion goes linearly to zero.
We see that the speed of sound and in particular the anisotropy is fairly robust against changes in temperature
deep in the superfluid phase. When the temperature approaches the critical temperature 
(in Figure \ref{fig:temperature} $T_c=0.15J$ (above) and $T_c=0.12J$ (below))
the speed of sound increases. This effect arises from the
temperature dependence of the gap $\Delta$ which falls to zero when $T\rightarrow T_c$.

\begin{figure}
\includegraphics[width=0.5\textwidth]{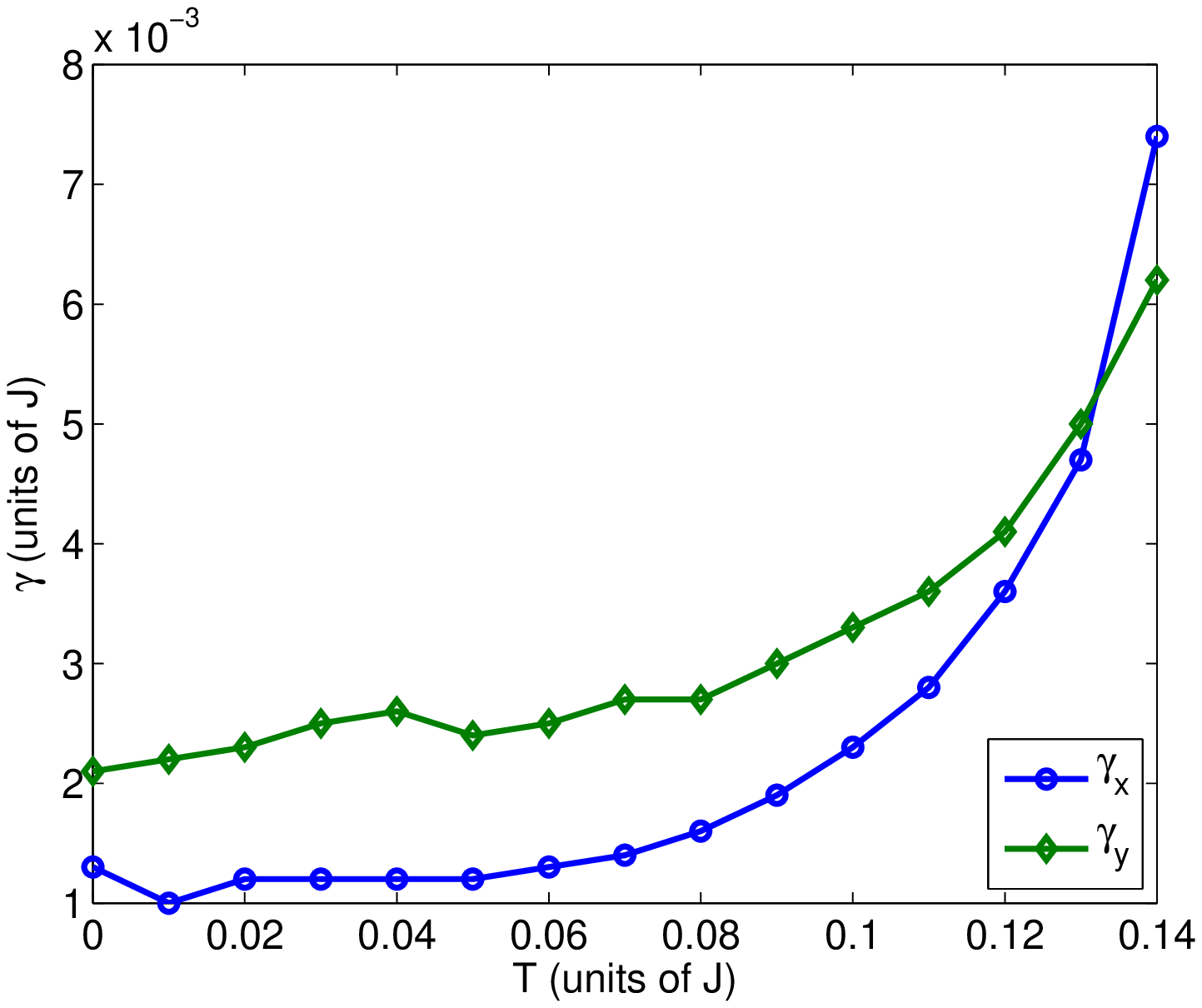}
\includegraphics[width=0.5\textwidth]{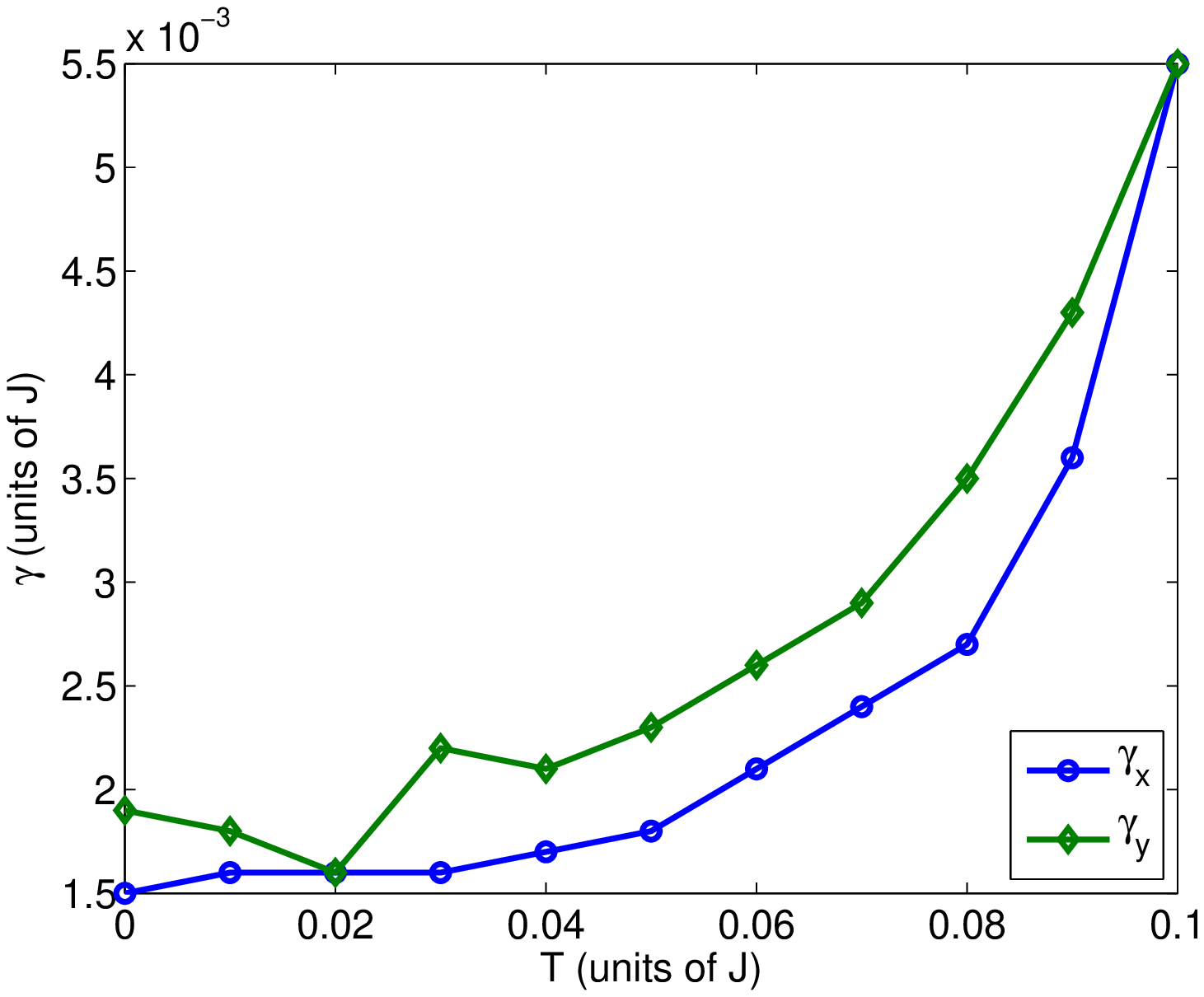}
\caption{(Color online) 
The damping rate $\gamma$ of the lowest $\kk$-mode, $k=2\pi/(dN_x)$ in directions
parallel ($\gamma_x$, circles) and perpendicular ($\gamma_y$, diamonds) to the FFLO vector $\q=q\evec_x$ 
as a function of temperature.
The parameters for the plots above and below are the same as in figure \ref{fig:temperature}.
}
\label{fig:damping} 
\end{figure}

In Figure \ref{fig:damping} we plot the damping rates of the $k=2\pi/(dN_x)$ collective mode. 
This is the lowest non-zero wave vector. 
As mentioned previously, in the FFLO state finite damping is present
already at $T=0$ in contrast to a BCS state, since the FFLO state has quasiparticle excitations even at $T=0.$
Increasing the temperature creates more quasiparticle excitations and the damping becomes stronger. 
The result implies that in a broad range of temperatures a significant portion of the damping can be attributed 
to the zero temperature quasiparticles as opposed to thermal excitations.
The damping is notably enhanced close to $T_c$ as the excitation gap $\Delta$ decreases rapidly close to $T_c$. 
The largest damping rates in the Figure \ref{fig:damping} are $13\%$ (above) and $15\%$ (below) of the 
corresponding collective mode frequency, 
which means that the low frequency collective modes are well-defined in the entire temperature range shown. 

\subsection{Quasi-1D lattice \label{sec_results2}}

\begin{figure}
\includegraphics[width=0.5\textwidth]{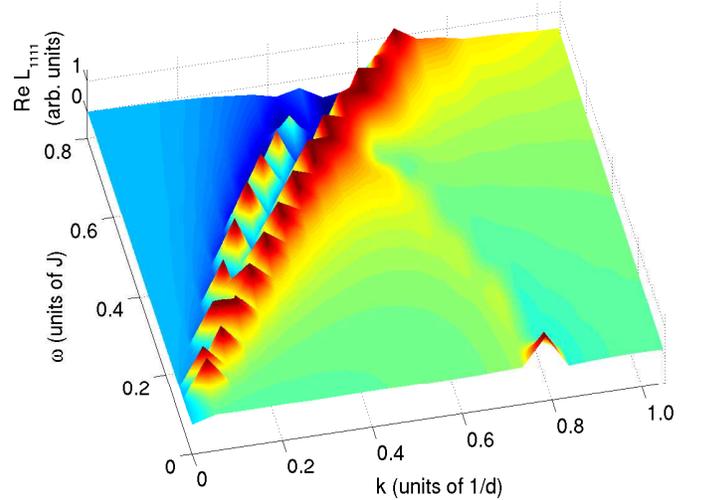}
\caption{(Color online) 
The real part of the density response function $L_{1111}(\kk,\omega)$ 
for an FFLO gas in a quasi-1D optical lattice along the $x$-axis.  
The hopping in the $y$ and $z$ directions is $J_y=J_z=0.02J_x$ and moreover
$\mu_1= 2.6 J_x$, $\mu_2= 1.4 J_x$, $T= 0.05 J_x$, $U= -2.7 J_x$, $\Delta= 0.3 J_x$ and $q= 8\pi/(dN_x)$
with $\q$ parallel to the $x$-axis.
There are two strongly peaked responses.
The lower branch corresponds to the phonon mode of the previous section while the upper branch is caused
by quasiparticle excitations centered into a narrow stripe due to the quasi-1D nature of the system.
}
\label{fig:quasi_1D} 
\end{figure}  \begin{figure}
\includegraphics[width=0.5\textwidth]{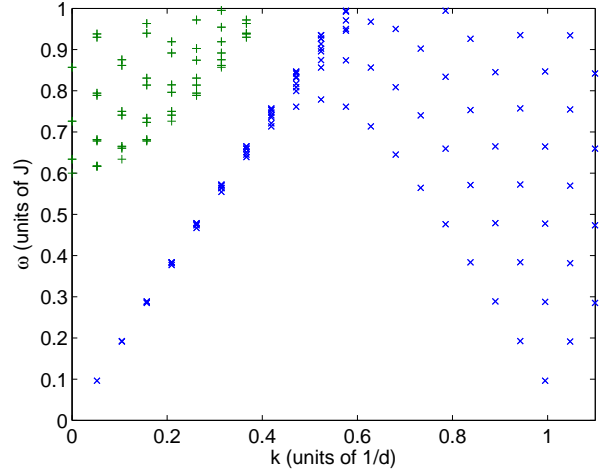}
\caption{(Color online) 
The zero temperature quasiparticle channels (for quasiparticles with zero transverse momentum) 
calculated for the system of Figure \ref{fig:quasi_1D}. 
A cross denotes a transition of a single quasiparticle 
and a plus a pair breaking which involves the creation of two quasiparticles.
At low wave vectors there is only a narrow
stripe of quasiparticle channels which becomes clearly visible also in the GRPA response.
}
\label{fig:quasi_1D_quasip} 
\end{figure}  

We now turn to quasi-1D optical lattices motivated by recent theoretical works which
predict that the FFLO state would be more stable in such geometries  \cite{orso2007,hu2007,parish2007,zhao2008,donghee}.
We study a three dimensional cubic lattice with lattice constant $d$ and a weak hopping in two directions \textit{i.e.}
$J_y=J_z\ll J_x$. The FFLO $\q$-vector is directed along the $x$-axis.

In the quasi-1D setup the collective mode spectrum along the different axes scales
with the appropriate hopping strength. Therefore, one observes that $c_y/c_x \sim J_y/J_x \ll 1$.
Now, it is still true that the FFLO state creates an additional difference perpendicular
to the $\q$-vector but the relative variation in $c_y$ and $c_z$ is in the same order of magnitude as in the 
case of an isotropic lattice and therefore negligible in comparison to the anisotropy caused 
directly by the hopping anistropy. Therefore, the analysis of the sound propagation in different directions
does not suggest direct means for observing the FFLO state in a single experiment
in systems that are strongly anisotropic already by geometry.

However, there is another interesting feature we find at the quasi-1D limit.
In the following we have an optical lattice with $N_L=120^3$ lattice sites 
\textit{i.e.} $N_x=N_y=N_z=120$ lattice sites in each direction,
and $J_y=J_z=0.02 J_x$.
In Figure \ref{fig:quasi_1D} we plot the density response function $L_{1111}(\kk,\omega)$ for an FFLO state
with $\mu_1= 2.6 J_x$, $\mu_2= 1.4 J_x$, $T= 0.05 J_x$, $U= -2.7 J_x$, $\Delta= 0.3 J_x$ and $q= 8\pi/(dN_x)$.
We now find two strongly peaked responses. Both of these branches are linear for small wave vectors and also gapless.
Moreover, the higher branch vanishes for $kd\approx 0.5$.
The lower branch corresponds to the collective density mode of a multidimensional lattice. 
The higher branch in turn is created by the FFLO quasiparticle transitions which are illustrated in Figure \ref{fig:quasi_1D_quasip}.
This higher branch is present also in $L_{1212}$ but with the opposite sign. Therfore it tends to cancel, though not exactly, 
the contribution of $L_{1111}$ in $\chi_1=L_{1111}+L_{1212}$. To illustrate the effect clearly we thus plot in this section $L_{1111}$.

In the quasi-1D limit the quasiparticle energies are dominated by the $k_x$ wave vector and the transverse momentum
only creates a relative variation on the order of $J_y/J_x$ to these energies.
This 1D-like dispersion of the quasiparticles places a notable restriction to the possible energy and momentum trasfers
for small energies and momenta. On the other hand, each possible quasiparticle transition gains a significant weight for the very same
reason by the following argument. Let $\delta E_0(k_x)$ be the energy of transition between an empty and a filled quasiparticle state
with no transverse momentum. This means that there are filled and empty
quasiparticle states at energies $E_{\pm}(k_x+p_x,p_y=0,p_z=0)$ and $E_{\pm}(p_x,p_y=0,p_z=0)$ for which 
$\delta E_0(k_x)=E_{\pm}(k_x+p_x,0,0)-E_{\pm}(p_x,0,0)$. Denoting then $\delta E(k_x)=E_{\pm}(k_x+p_x,p_y,p_z)-E_{\pm}(p_x,p_y,p_z)$  
we find after some straightforward algebra that $\delta E(k_x)-\delta E_0(k_x) \sim J_y k_xd$, for small $k_x$. In other words 
all the quasiparticles with different momenta $p_y$ and $p_z$ but the same momentum $p_x$ contribute at a narrow energy interval
on the order of $J_y k_xd$ and therefore we find a strongly peaked response in Figure \ref{fig:quasi_1D} following the
quasiparticle energies of Figure \ref{fig:quasi_1D_quasip}. The strong quasiparticle response vanishes towards higher wave vectors as the
quasiparticle transitions are spread to a wide energy range. 
Moreover, for low wave vectors the dispersion of the lower collective mode
does not overlap with the quasiparticle excitations and therefore the mode is undamped,
unlike in the higher dimensional case.

To compare to the results of \cite{edge2009} for a strict 1D system and an LO ansatz
(cosine form order parameter), it is interesting to notice 
that such a system has a similar two mode structure at low 
wave lengths as the FF ansatz (plane wave order parameter) 
in a quasi-1D setup. However, at higher wave lengths 
the mode associated with the excess quasiparticles 
does not exhibit an additional Brillouin zone
structure in the FF case. The reason is that this Brillouin zone 
structure is caused in the LO case by the oscillatory structure of the 
quasiparticle density profile \cite{edge2009}, while the FF quasiparticle density
profile is uniform.

\section{Conclusions \label{sec_conc}}

We have studied the density response of
a spin-imbalanced ultracold Fermi gas in an optical lattice in the FFLO state.
Using the Kadanoff-Baym formalism we derived the linear response function for this system in the
generalised random phase approximation. 
We then calculated the collective mode spectrum in a 2D square optical lattice
and showed that the speed of sound is anisotropic due to the anisotropy of the FFLO pairing.
This suggests an experiment in which one monitors the propagation of a local density perturbation 
in order to find evidence of the anisotropic pairing mechanism of the FFLO state.

Moreover, we studied the damping of the collective modes and showed that despite the presence of quasiparticles
in the FFLO ground state the collective modes have a relatively weak damping rate and are thus well-defined and
physically meaningful elementary excitations of the system.

We also studied a quasi-1D system. In this case the anisotropy of the sound propagation is predominantly
caused by the anisotropy of the lattice itself in contrast to a possible exotic pairing mechanism.
However, the quasi-1D system is qualitatively different from higher dimensional systems as it contains
an additional collective-type response of quasiparticles.

To draw future quidelines, a clear way to improve on the
results presented in this paper would be the inclusion 
of more elaborate order parameter structure,
in particular the LO ansatz with a cosine type order parameter. 
In this case, one cannot simplify the problem in momentum
space to the same extent as in section \ref{sec_response2} of this paper, 
and a heavier numerical method in position space is called upon.
One would still anticipate an anisotropic speed of sound for the LO ansatz as well,
based on presenting the state as a superposition of two FF states, 
as well as by the arguments of \cite{radzihovsky2011}.

\begin{acknowledgments}
We thank D.-H.\ Kim, J.J.\ Kinnunen and A.\ Korolyuk for useful discussions. 
This work was supported by Finnish Doctoral Training Programme in Computational Sciences, 
EUROQUAM/FerMix and Academy of Finland 
(Project No. 210953, No. 213362, No. 217043, No. 217045, No.135000, No. 141039),
and conducted as a part of a EURYI scheme grant, 
see www.esf.org/euryi.  
\end{acknowledgments}


\end{document}